\documentclass[%
 aip,
 jmp,%
 amsmath,amssymb,
 reprint,%
]{revtex4-2}

\usepackage{graphicx}
\usepackage{dcolumn}
\usepackage{bm}

\usepackage[utf8]{inputenc}
\usepackage[T1]{fontenc}
\usepackage{mathptmx,color,soul}

\begin{document}

\preprint{AIP/123-QED}

\title{Non-destructive detection of large molecules without mass limitation} 



\author{A. Poindron}
\author{J. Pedregosa-Gutierrez}
\author{C. Jouvet}
\author{M. Knoop}
\author{C. Champenois}
\email[corresponding author : ]{caroline.champenois@univ-amu.fr}
\affiliation{Aix Marseille Univ, CNRS, PIIM, Marseille, France}


\date{\today}

\begin{abstract}
The problem for molecular identification knows many solutions which include mass spectrometers whose mass sensitivity depends on the performance of the detector involved. The purpose of this article is to show by means of molecular dynamics simulations, how a laser-cooled ion cloud, confined in a linear radio-frequency trap, can reach the ultimate sensitivity providing the detection of individual charged heavy molecular ions. In our simulations, we model the laser-cooled Ca$^+$ ions as  two-level atoms, confined thanks to a set of constant and time oscillating electrical fields. A singly-charged molecular ion with a mass of  $10^6$ amu is propelled through the ion cloud. The induced change in the fluorescence rate of the latter is used as the detection signal. We show that this signal is due to a significant temperature variation triggered by the Coulombian repulsion and amplified by the radio-frequency heating induced by the trap itself. We identify the optimum initial energy for the molecular ion to be detected and furthermore, we characterize the performance of the detector for a large range of confinement voltages.
\end{abstract}

\pacs{keywords: mass spectrometry, non-destructive detection, giant molecules}

\maketitle 

\section{Introduction}
Mass spectrometry is among the most advanced techniques of precise identification and fingerprinting today and covers a broad range of species from light atoms up to giant molecules, like proteins or viruses. This method can reach extreme resolution by exploiting various techniques \cite{shinholt_frequency_2014,keifer_charge_2014} and in particular if coupled to sophisticated time-of-flight trajectories \cite{blaum2006,toyoda_multi-turn_2003}. The vast majority of mass spectrometers identify the species by its mass-to-charge ratio which can be done by measuring either a characteristic frequency, arrival time, or any  properties reflecting the ion trajectory within electromagnetic fields.  For these mass spectrometry measurements, the sample species has to be ionized, which can be done in different configurations (ESI, MALDI, electron impact).  It will then undergo a mass-filtering process in order to separate different species (TOF, quadrupole filter, traps, ...), before finally reaching a detector. The need to expand the sensitivity range of charged particle detectors  towards  very large mass has appeared already in the 90's \cite{frank99} with the achievement of heavy molecules ionising sources like MALDI and ESI.

In mass spectrometry, the detection  is most often done by accelerating the species under identification (SUI) towards a charge detecting device (e.g. an electron multiplier, a micro channel plate (MCP), ...)  \cite{contino_charge_2013,knoop_detection_2014}. Such detectors are based on secondary electron emission: the incident particle  must generate the emission of at least one electron. This electron is then amplified in a cascade process, in order to generate a measurable current at the final anode. For increasing molecular mass, triggering the first electron by impact becomes less likely as velocities are getting smaller for a given energy value\cite{twerenbold01,takahashi11}. The low detection efficiency of electron multipliers and MCPs is known and understood \cite{gilmore00,fraser02} for the mass range beyond 10$^4$ a.m.u  and the strong dependence of the detection efficiency with the incident ion mass, for a given impact energy, is an issue for a mass spectrometer. If not corrected by further amplification, this variable efficiency produces a mass spectrum which does not accurately reflect the mass abundancy of the incoming ions. This drawback is circumvented with calorimetric cryo-detectors  whose application range now covers  masses as large as few MDa when coupled to TOF devices \cite{twerenbold96b,twerenbold96,frank99} and which show a high detection efficiency independent of the mass \cite{twerenbold01,keifer17} but  require a cryogenic environment.

Interest in measuring accurate masses for species with molecular weights much greater than 1~MDa \cite{frank99,tito00} has led to the development of  single-particle techniques \cite{keifer17}, like shown by the recent detection of individual ions carrying a single charge by charge detection mass spectrometry \cite{todd20}. In this manuscript we propose an original detection method with the potential to non-destructively detect single molecules without limitation of the mass range. Our novel approach consists in a radical change in the detection principle of the molecular ion, based on the perturbation that it induces in crossing a laser-cooled cloud of trapped ions. The SUI deposits part of its kinetic energy in  the trapped ion cloud and the induced temperature increase  is amplified by the radio-frequency (RF) heating, characteristic of  electrodynamic traps. The exploited signal is the corresponding change in the laser induced fluorescence of the laser-cooled stored ion cloud, which can be tuned to be sufficiently long-lasting to be observable. As the ion cloud is just heated and not lost, it can easily be reset to initial (cold) conditions by the tuning of the laser according to a cooling protocol. Therefore, this method provides a non-destructive detection system without mass limitation, neither charge number requirement, for individual molecules \cite{pat_giantmol}.

This article presents the operation principle of this novel detector,  demonstrated by means of molecular dynamic simulations, which allow to scan a vast range of parameters. Section II  describes the scheme of the detection process, and introduces its main properties. In section III, we then describe the details of the simulation, which takes up the realistic environment and follows the injected species throughout the detection cloud. Section \ref{sec:full} describes a full sequence of the detection process for one particular set of parameter. The best choice for the SUI initial energy and the cloud trapping parameters are discussed in section \ref{sec:efficiency}. The experiment under construction will be briefly outlined in the conclusions.

\section{Working principle of the detector and of its simulation}

\begin{figure}[h!]
 \centering
 \includegraphics[width=.35\textwidth]{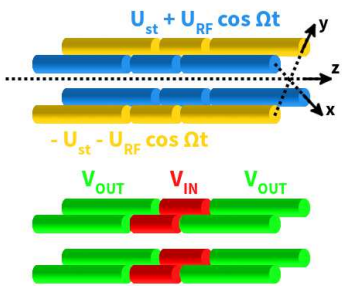}
 \caption{Schematic of the rf trap and the voltages applied to each electrode. The inner radius of the trap is $r_0$=2,5~mm and the central electrodes are 4~mm long. }
 \label{fig:trap_schema} 
\end{figure}

The key element of the detector is a cloud of laser-cooled atomic ions stored in a linear RF quadrupole trap, very similar to many other experiments \cite{pedregosa10a,pedregosa15}. Ca$^{+}$ ions have been chosen due to the commercial availability of lasers at the required wavelengths for photoionisation \cite{kjaergaard00} and laser-cooling, as well as for the observation efficiency \cite{knoop04}. These ions (with mass $m$) are trapped in a linear RF trap (inner radius $r_0$=2,5~mm) as depicted in figure~\ref{fig:trap_schema}, built and connected as the one in reference\cite{herskind09}, which leaves the trap's $z$-axis free of electrodes and allows easy injection. The radial trapping potential is given by:
\begin{equation}
\Phi (x,y,t) =\left(U_{st} + U_{RF} \cos{\Omega t}\right) \frac{(x^2 - y^2)}{r_0^2} 
\end{equation}
where $U_{st}$  is a the static voltage and $U_{RF}$ is the amplitude of the oscillating voltage applied to the rods, see figure~\ref{fig:trap_schema}. The potential along the trap axis, $z$, can be considered as harmonic on the length scale of the ion cloud and is well represented by $U_H(z) = m \omega^2_z z^2/2$ with $\omega_z^2$ scaling linearly with the voltage difference $U_{DC}=V_{out}-V_{in}$ (see figure~\ref{fig:trap_schema} for the voltage definition).

A  cloud of 1024 Ca$^+$-ions is laser-cooled to a temperature below 10~mK, and its fluorescence is monitored with the cooling laser set at fixed frequency.  A heavy molecule with mass $m_{SUI} = 10^6$~amu and a unit charge $Q=+e$  is  injected into the trap, at a distance larger than 1~mm  from the ion cloud  with an initial position exactly on the trap symmetry axis and a given initial kinetic energy.  The injected molecule is heavy enough not to be deviated by the trap potential nor by the interaction with the ion cloud. In our simulation we can extract its energy loss, an important parameter in the understanding of the energy exchange between the injected molecule and trapped ion cloud. The equilibrium state of the  laser-cooled ion cloud is perturbed by the molecule crossing and the kinetic energy of the cloud increases. In a second step, the heating process  is amplified by radio-frequency heating \cite{dehmelt67}.  The temperature of the cloud is not accessible to measurements and its indirect observation is based on the collection of the laser induced fluorescence. It is numerically computed by the number of photons emitted by the ion cloud in a time cell duration $\Delta t$. At fixed laser frequency $\omega_L$, this number is sensitive to the ion velocity $\vec{v}_i$ through the Doppler effect which shifts the laser frequency seen by the ion to $\omega_L-\vec{k_L}.\vec{v}_i$ with $\vec{k_L}$ the laser wave-vector. The probability for ion $i$ to be in the excited state is then
\begin{equation}
\label{eq:P_exc}
P_{e} (i)=\frac{\Omega_r^2 / 4}{ (\delta - \vec{k_L}.\vec{v}_i)^2 + \Gamma^2/4 + \Omega^2_r/2 }
\end{equation}
where $\Omega_r$ is the Rabi frequency standing for the atom-laser coupling strength, $\Gamma$ is the spontaneous emission rate of the excited level and $\delta$ is the laser detuning $\omega_L-\omega_0$ with respect to the atomic transition frequency, for an atom at rest. On average, the number of photons emitted by an ion during $\Delta t$ is $P_e(i)\times \Gamma \times \Delta t$.

To have a better understanding of the phenomena that are discussed in the following, some features concerning the self-organization of a laser-cooled ion cloud are useful. They can be demonstrated in the static picture where the cloud can be considered as trapped by a static pseudo-potential\cite{dehmelt67} which is the sum of the harmonic approximation of the DC potential designed to trap along the axis $z$, $U_H(z)$ and a radial potential which is also harmonic and can be written\cite{drewsen00,drakoudis06} as $U_{pp}(r) = m \omega^2_r (x^2+y^2)/2$ with $\omega_r^2=\omega_x^2-\omega_z^2/2$ and $\omega_x=e\sqrt{2}U_{RF} /(m r_0^2 \Omega^2)$ with $e$ the elementary charge. Rigorously, this static picture is a close description of the RF-trapping only for $U_{RF}$ values lower than the ones used for the simulations presented here. Nevertheless, it gives useful clues to understand the impact of the trapping parameters.  In the pseudo-potential picture, it has been shown in references\cite{turner87,hornekaer02} that a cold ion cloud has  a uniform density over the whole sample and forms a spheroid  with radius $R$ and half-length $L$, with an aspect ratio  $\alpha=R/L$ that depends on the aspect ratio of the 3D-potential $\omega^2_z/\omega^2_r$. For prolate clouds like used in the present simulations, the relation is\cite{turner87} 
\begin{equation}\label{eq:aspect}
\frac{\omega^2_z}{\omega^2_r} = -2\frac{\sinh^{-1}(\alpha^{-2}-1)^{1/2} - \alpha(\alpha^{-2}-1)^{1/2}}{\sinh^{-1}(\alpha^{-2}-1)^{1/2} - \alpha^{-1}(\alpha^{-2}-1)^{1/2}}
\end{equation}
which can be simplified in $\omega^2_z/\omega^2_r \simeq 2 \alpha^2 \ln(1/\alpha)$ for $\alpha<0.3$.

When $U_{RF}$ and/or $U_{DC}$ are changed, the shape of the cloud is modified according to Eq.~(\ref{eq:aspect}) but one can show\cite{champenois09} that the cloud density depends only on $U_{RF}$ and scales like $U^2_{RF}$. The next section describes the numerical simulations built to produce a signal useful to evaluate the detection efficiency of this cloud.

\section{Molecular Dynamic Simulations}\label{sec:MDsimu}

The molecular dynamics simulations presented in the following numerically  integrate the equations of motion of $N$ interacting ions within  the trapping potential oscillating at frequency $\Omega / 2\pi = 2$~MHz and with no static contribution ($U_{st}=0$).  The time step of the integration is chosen to be $2\pi/\Omega /100= 5$~ns. All the relevant parameters used in these simulations are gathered in the appendix.

A complete simulation consists of 5 steps, with each one having its own specifically designed code :
\begin{enumerate}
\item initialization and thermalization  to a predetermined temperature of the  ions trapped by means of a low value of $U_{RF}$,
\item the thermalization process is replaced by laser-cooling, modeled by the recoil induced by absorbed and emitted photons,
\item linear increase of the RF voltage until the desired $U_{RF}$ is reached keeping the same laser-cooling protocol,
\item injection of the SUI by setting it on the trap axis, off the ion cloud, with a given energy,
\item observation of the post-crossing dynamics of the ion cloud, calculation of the cloud temperature evolution and record of the number of emitted photons.
\end{enumerate}

In the first step, ions created with a null velocity and a position randomly chosen in a Gaussian distribution evolves in a low-RF trapping field ($U_{RF}$=26.9~V, corresponding to a Mathieu parameter\cite{Major05} $q_x=0.25$). Their thermalization is  modeled by a Langevin process involving a friction term while the heating term is taken into account as a single thermal bath. Such a problem is described by the following set of Langevin  equations which adds to the Coulomb repulsion between the ions: 
\begin{eqnarray}
m \partial_{tt} x_i &=& Q^2 k_C\sum_{j=1,j\neq i}^N\left({\frac{x_i - x_{j}}{|\vec{r}_{i}-\vec{r}_{j}|^3}} \right)- \frac{2 Q U_{RF} \cos{\Omega t}}{r_0^2} x_j- \gamma\partial_t x_i + \sqrt{2\Gamma k_B T_b} \theta_{xi} \nonumber\\
m \partial_{tt} y_i &=& Q^2 k_C\sum_{j=1,j\neq i}^N\left({\frac{y_i - y_{j}}{|\vec{r}_{i}-\vec{r}_{j}|^3}} \right) + \frac{2 Q U_{RF} \cos{\Omega t}}{r_0^2} y_j- \gamma\partial_t y_i + \sqrt{2\Gamma k_B T_b} \theta_{yi} \nonumber \\
m \partial_{tt} z_i &=& Q^2 k_C\sum_{j=1,j\neq i}^N \left({\frac{z_i - z_{j}}{|\vec{r}_{i}-\vec{r}_{j}|^3}} \right) - \left|\frac{d U_G(z)}{d z}\right|_{z_i} - \gamma\partial_t z_i +\sqrt{2\Gamma k_B T_b} \theta_{zi},
\end{eqnarray}
where $\vec{r} = (x,y,z)$, $k_C=1/(4\pi\epsilon_0)$, $U_G(z)$ is the axial potential given by Eq.~(\ref{eq:UG}), $\gamma$ a friction coefficient, $k_B$  Boltzmann's constant, $T_b$ is the temperature of the thermal bath  and $\theta_{xj}$, $\theta_{yj}$ and $\theta_{zj}$ are a collection of independent standard Wiener processes~\cite{skeel02}. The equations of motion in this first code are numerically solved using the vGB82 algorithm as described in reference~\cite{skeel02} and already implemented in reference~\cite{pedregosa20}.

The rest of the simulation codes do not include any friction term or thermal bath and use the Velocity-Verlet algorithm. Laser-cooling is introduced in step 2  by an algorithm \cite{marciante10} which uses a two-level atom model to compute the recoil induced by photon emission and absorption. When in the ground state,  the probability for an ion to absorb a photon depending on its instantaneous velocity $\vec{v}_i$, is based on Eq.~\ref{eq:P_exc}, and at each time step, it is compared is to a random number to decide if the ion is excited or not at the next time step. If excited, the ion velocity is modified according to the recoil of a  photon momentum $\hbar \vec{k}_L$. Once the ion is classified as excited, the algorithm computes the probability for spontaneous and stimulated emission and compares them to  another random number to decide if the ion emits a photon. In the case of spontaneous emission, the photon emission direction is random with an isotropic probability in space. Such an approach allows to model a realistic laser cooled ion cloud upon which the external particle can be injected. Moreover, by using a photon absorption / emission approach, it is possible to record the total number of emitted photons which represents one of the few quantities accessible to measurement in a real experiment. Step 3 continues with the same processes but the radio-frequency voltage amplitude $U_{RF}$  is increased to reach the value chosen for the SUI-cloud interaction step.

At the beginning of step 4, the SUI is initialized on the trap axis at $x=y=0$ and $z = -z_i=-1.5$~mm with an initial energy, $E_{SUI}$. The variation of the electrostatic potential along $z$ drives the SUI to the ion cloud and the code stops when the SUI reaches $z = +z_i$, with $z=0$ the trap center. The description of the potential along the trap axis as harmonic is sufficient on the cloud size scale but it does not fit with the long distance profile of the potential which flattens out of the central part of the trap (see Fig.~\ref{fig:trap_schema}). This correction must be taken into account in the code for  the initialization of the SUI at position $z_i$, to make possible situations where the chosen initial energy $E_{SUI}$ is lower than  $ m \omega_z^2 z_i^2/2 $. The potential equation is adapted for long distance $|z|$ to the trap center by fitting the axial potential generated by the trap geometry and calculated by the commercial finite element method software SIMION8.0  \cite{simion}. This leads to the equation
\begin{equation}
\label{eq:UG}
U_G(z) = m \omega_z^2  L^2 (1-e^{-z^2 / 2 L^2 })
\end{equation}
 where $L=2.45$~mm is the relevant length as given by the fit which behaves like a harmonic potential at its bottom, characterised by $\omega_z / 2\pi = 90.8$~kHz for  $U_{DC}=1$~V, also deduced from the fit.

 While the SUI is present in the simulation, the time integration concerning the dynamics driven by the Coulomb interaction and the RF trapping is run with a variable time step Velocity-Verlet algorithm adapted to the closest distance between the charged particles, to make sure  the interaction is properly described \cite{sillitoe17}. This variable time step procedure demands extra computations and therefore is limited to step 4. The algorithm  describing the laser-ion interaction keeps the same constant time step of 5~ns and with an  excited state lifetime of 7~ns for the chosen Ca$^+$-ion, it is justified to use the stationary limit of Eq.~\ref{eq:P_exc}   to compute the fluorescence for each ion, even if its velocity changes in time \cite{champenois16}. To avoid irrelevant statistical fluctuations, the number of emitted photons is recorded on time bins of size $\Delta t=1 \mu$s.

\section{Results of a full detection simulation}\label{sec:full}

\begin{figure}
 \centering
 \includegraphics[width=13.cm]{ 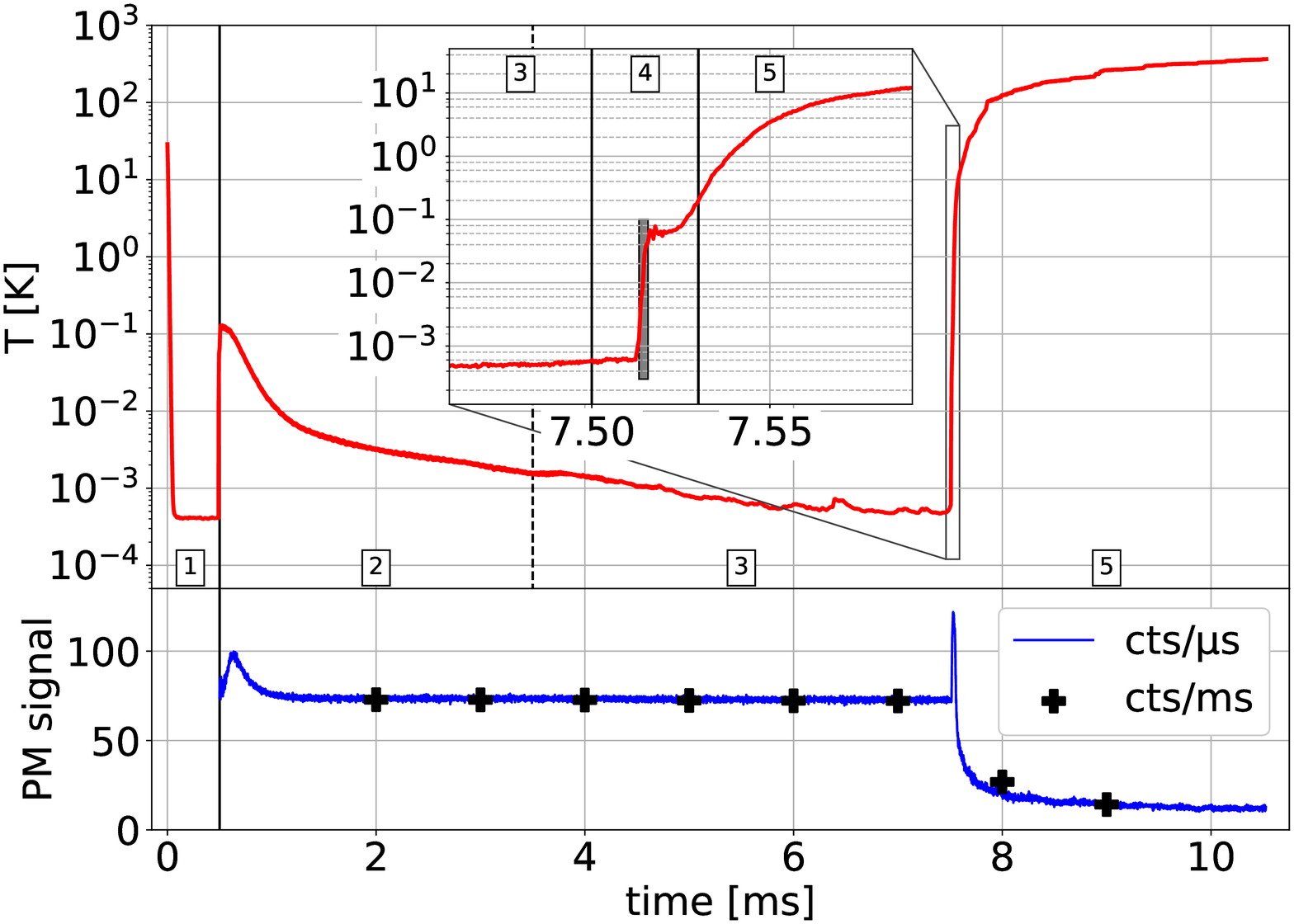}
 \caption{Results of the simulation with parameters given in section \ref{sec:full}. Top : temperature of the cloud during one run. The framed numbers indicate which part of the simulation is being executed according to description of section \ref{sec:MDsimu}. The inset focuses on the fourth part where the SUI is injected. The grey rectangle indicates when the SUI is actually crossing the ion cloud. Bottom : ion cloud fluorescence. The blue line represents all the photons emitted by the cloud integrated during 1~$\mu$s. The black marks represent the fluorescence signal ${S}$ as it would be measured with realistic conditions : a 1~ms integration time and a $10^{-3}$ detection efficiency. }
 \label{fig:example_full_simu}
\end{figure}

Figure~\ref{fig:example_full_simu} shows the result for a complete simulation run for $N=1024$ Ca$^+$ ions. The trap parameters for this case are $U_{RF} = 64.6$~V, $U_{st}=0$ and $U_{DC}=3$~V which correspond to a Mathieu parameter \cite{Major05} $q_x=0.6$ and $\omega_z / 2\pi = 157.3$~kHz. The SUI  has a mass  $m_{SUI} = 10^6$~amu, a unit charge and its initial energy $E_{SUI}$  is such that its kinetic energy at the minimum of the trapping potential well $E_0$ is equal to 50~eV. The top part of Fig.~\ref{fig:example_full_simu}  shows the time evolution of the ion cloud temperature while the bottom part traces the fluorescence signal, with two different average time scales. The temperature of the cloud is computed from the sample averaged squared velocities,  averaged over one RF-period. This strategy to compute time-averaged values is used to eliminate the RF-driven motion contribution from the velocity \cite{prestage91b, schiffer00,marciante10}. The number of photons emitted by the entire cloud is given for a 1~$\mu$s acquisition time, which is a time scale relevant for the simulation. The signal $S$ expected from a realistic experiment is also plotted, assuming a detection efficiency of $10^{-3}$ and an acquisition time of 1~ms, which is typical for this kind of experiment, to reach a relevant signal to noise ratio (SNR) equal to $1/\sqrt{S}$ when taking into account the photon counting noise.

The first step, from $t=0$ until  $t=0.5$~ms,   corresponds to the preparation of the cloud in the radio-frequency trap with a low RF voltage, which brings it in equilibrium with a thermal bath temperature. The value of the thermal bath temperature, $T_{b} = 0.5$~mK, is chosen equal to the Ca$^{+}$ limit temperature of the Doppler laser cooling. At such temperature, the ion cloud forms a Coulomb crystal \cite{drewsen98} with the shape of a spheroid with a length of 470~$\mu$m and a radius of 47~$\mu$m. During the second step, covering $t=0.5$~ms to $t=3.5$~ms, the simulated ions are closer to the experimental situation as they are now submitted to the laser-cooling Velocity-Verlet algorithm, with a laser beam propagating along the trap axis $z$. The laser detuning used for all the results presented in this paper is $\delta=-\Gamma$. The choice for this laser detuning  results from a compromise between cooling efficiency and signal discrimination as it controls the velocity class with the highest probability of excitation and the fluorescence rate for a  given velocity distribution. With the typical temperatures of a large set of simulations, a detuning of $\delta=-\Gamma$ appears to be the best compromise to keep the ion cloud cold before the SUI crossing and reach a large fluorescence signal difference when the SUI has left. The choice for the coupling strength $\Omega_r$ is governed by a compromise between the temperature that can be reached by laser cooling and the photon diffusion rate \cite{lett89}. For all the simulations discussed in this paper, it is chosen such that $\Omega_r=\Gamma$.

At the end of step 3, the ion cloud has reached a stationary state in the  RF-trapping field defined by $U_{RF} = 64.6$~V (Mathieu parameter $q_x=0.6$). The creation and disappearance of the SUI at position $-z_i=1.5$~mm and $+z_i$ from the trapping well minimum are the beginning and end of step 4 and are materialised by a vertical line on the  inset of Fig.~\ref{fig:example_full_simu}, which shows a zoom into the temperature evolution of the ion cloud for steps 4 and 5. This value was chosen far enough from the trap center so that the initialisation of the SUI does not perturb the trapped ion cloud. The first sharp jump of the temperature around $t \approx 7.52$~ms is a signature of the passage of the SUI through the ion cloud. The time actually spent by the SUI inside the ion cloud is materialised by a grey shadow on the results and lasts 2.5~$\mu$s. The following simulation sequence represents the evolution of the ion cloud under laser cooling after the passage of the injected SUI.

 Following the SUI injection, the fluorescence rate  shows a  short-lived  increase  during and shortly after the passage of the projectile followed by a significant drop to a stationary value which is 6 times smaller than the value it had before the molecule passage. This behavour is different from the temperature profile which increases in two steps :  starting from the initial equilibrium value close to 0.5~mK, a first jump is observed to values of the order of few tens of millikelvin, as a direct consequence of the perturbation caused.  Then, once the projectile has left the cloud, the temperature keeps increasing to reach values of the order of several hundreds of Kelvin.  The correlation between temperature and fluorescence rate sits in the Doppler effect which impacts the laser induced excitation probability.  The plot of the  fluorescence signal ${S}$ on Fig.~\ref{fig:example_full_simu} shows a short lived increase that happens when the cloud reaches a temperature of the order of 1~K for which the fluorescence rate is maximal for the chosen laser detuning. As the cloud temperature keeps increasing, the fluorescence drops on a time scale shorter than 0.1~ms and the short lived increase cannot be observed in typical experimental conditions where a photon detection efficiency of $10^{-3}$ is assumed.  Therefore, a successful detection requires that the difference between the  stationary signal before and after the SUI crossing is larger than the detection noise and is persistent for a few ms. Like shown by the comparison of the top and bottom curves of figure~\ref{fig:example_full_simu}, the long term decrease of the fluorescence is correlated with the increase of the cloud temperature to several hundreds of Kelvin. This post-crossing temperature increase is due to RF-heating, a  side-effect of RF-trapping when trapped particles collide with each other or with a background gas \cite{dehmelt67} or when non-linear resonances occur between the rf-frequency and the oscillation frequencies of the ion in the trap\cite{alheit95, alheit96,drakoudis06}. Here, the  simulated collisions are the Coulomb collisions between the projectile and each ion of the target cloud, as well as the Coulomb collisions between the ions of the target. The two-step evolution of the temperature can be understood as first, an energy increase induced by the energy lost by the SUI inside the cloud, followed by a larger energy increase induced by RF-heating.

For an efficient detection, the energy loss of the SUI must be sufficient to trigger a perturbation of the cloud large enough to ignite the increase of the RF-heating rate. In the present simulation, the energy lost by the SUI is 11.48 meV. If we assume that this energy is transferred only to the thermal kinetic energy of the cloud, it would result in a temperature increase of 86~mK. The numerical simulations give a value close to 40~mK, showing that part of the lost energy is transferred also to the potential energy of the cloud, most probably the Coulomb interaction potential energy as ions have moved from their equilibrium position. 

The dependence of the heating rate with the  ion cloud temperature  studied numerically in \cite{ryjkov05} shows that, for given trapping parameters, it increases by several orders of magnitude when the temperature of the cloud increases from 0.1 to 1 K. This increased slope is also visible in the results of Fig.~\ref{fig:example_full_simu} for the same range of temperature.  If the heating rate in a perturbed cloud is higher than the laser cooling rate for the corresponding velocity distribution, the temperature of the cloud keeps increasing until it reaches a stationary value where the RF-heating rate is negligible because the ions do not interact strongly any more. The numerical results in \cite{ryjkov05,tarnas13} also show a strong dependence of the heating rate with the RF amplitude voltage $U_{RF}$. It is then possible to tune the RF-heating rate  to adapt it to the temperature of the cloud after the SUI crossing. The next section focusses on the dependence  of this initial energy deposition on the energy $E_0$ of the SUI and on the influence of the trapping parameters on the overall detection efficiency.

\section{Efficiency of the detector}\label{sec:efficiency}

\begin{figure}
 \centering
 \includegraphics[width=0.8\textwidth]{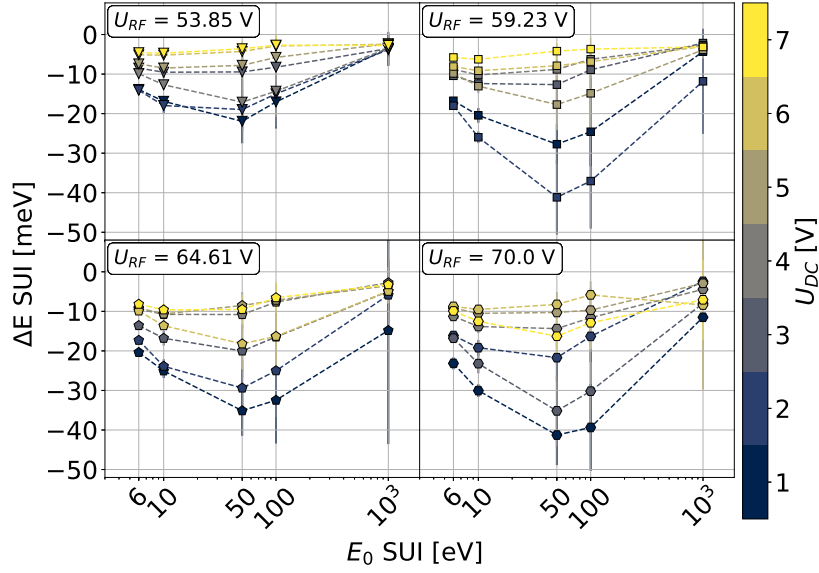}
 \caption{ Energy lost by the SUI $\Delta E$ versus the SUI kinetic energy $E_0$ at the center of the trap. 28 different sets of ($U_{RF}$, $U_{DC}$) are tested by 20 independent simulations. The average value is plotted with its error bar standing for $\pm 1$ standard deviation. Each of the four subplots stands for one $U_{RF}$, and the colour code translate the values of $U_{DC}$ (see the color chart on the right side of the figure).}
 \label{fig:delta_EGMol_allcases}
\end{figure}

The energy exchanged by a cold charged ensemble and a charged projectile has been studied  in the frame of the stopping power of a plasma\cite{zwicknagel_stopping_1999}. Simulation runs  for the  particular case of cold ions in a RF trap\cite{bussmann06b,sillitoe17} consider the energy transfer between the projectile and the trapped ion cloud with a detailed analysis of the projectile's energy evolution during its passage through the ion cloud. It has been shown that two different mechanisms contribute to the energy interchange: collective effects and Coulomb binary collisions and that the energy exchange rate depends on the kinetic energy of the projectile. An analytic estimation of the energy loss is extremely complex,  and the detailed study of the interaction between the injected particle and the trapped ion cloud goes beyond the aim of the present work. To estimate the most adapted  kinetic energy $E_0$  to be used for the injected particle, to favor a large  SUI energy loss $\Delta E$,  5 different kinetic energies $E_0$ have been studied: 1000~eV, 100 ~eV, 50~eV, 10~eV and 6~eV.  For different combinations of $U_{RF}$ and $U_{DC}$, the SUI is launched with an initial energy calculated such that its kinetic energy at the trapping potential minimum is $E_0$. The results are shown in figure~\ref{fig:delta_EGMol_allcases} where each data point represents the mean of 20 independent simulations. The large deviation observed on $\Delta E$ can be attributed to the contribution of binary collisions, that are very sensitive to  the exact position of the trapped ions at the moment of the passage of the particle. 

Some trends can be identified on Fig.~\ref{fig:delta_EGMol_allcases} : for the 4 different $U_{RF}$ values, $\Delta E$ increases when $U_{DC}$ decreases.  Furthermore, the results show that $\Delta E$ increases with $U_{RF}$. As these two modifications result in a longer cloud, we can conclude that there is a correlation between the length of the cloud and the energy loss. Depending on the ($U_{RF}$, $U_{DC}$) combinations, the variations of $\Delta E$ with $E_0$ are flat or show an extremum for $E_0=50$~eV. This value is chosen for the next simulations. 

\begin{figure}
 \centering
 \includegraphics[width=0.8\textwidth]{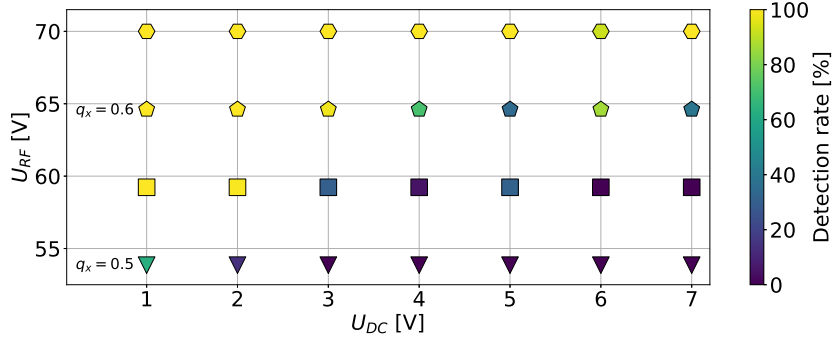}
 \caption{Detection rate as a function of $U_{RF}$ and $U_{DC}$ for $E_0 = 50$~eV. $q_x$ is the Mathieu parameter associated to the RF trapping. See the color chart that codes the detection rate. }
 \label{fig:main_results}
\end{figure}

To identify the conditions for an efficient detection,  we have performed simulations  over a range of values of $U_{RF}$ and $U_{DC}$, keeping the other simulations parameters  identical to those of figure~\ref{fig:example_full_simu}. The detection efficiency is defined by a criteria based on the signal-to-noise ratio SNR of the estimated experimental signal.
The detector is assumed to be efficient if the variation in the useful signal $S$ is larger than the statistical noise $\sqrt{S}$ of the signal before the SUI passage.
The detection rate,  shown on Fig.~\ref{fig:main_results},  indicates the probability of detection for 100 independent simulation runs . 

\begin{figure}
 \centering
 \includegraphics[width=0.8\textwidth]{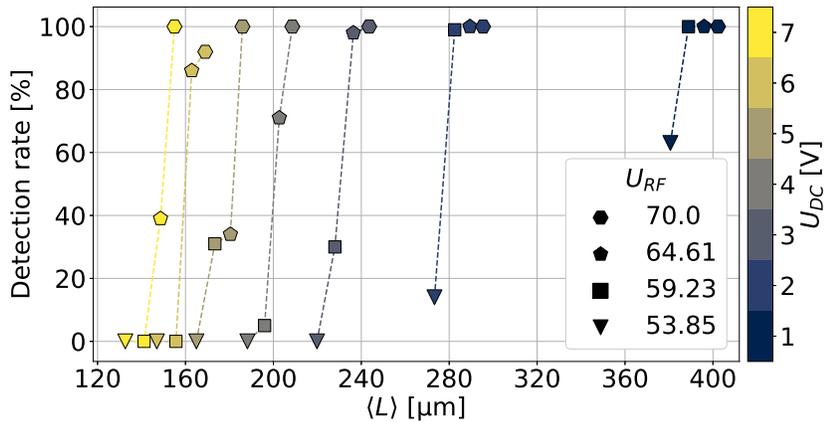}
 \caption{Detection rate as a function of the average cloud half-length $\left<L\right>$. The marker indicates the $U_{RF}$ and the color codes for the $U_{DC}$ values.}
 \label{fig:main_results_RandR/L}
\end{figure}

Figure~\ref{fig:main_results} shows that the detection rate increases with $U_{RF}$ and a 100\% efficiency is reached for the highest value of $U_{RF}$, for nearly all values of $U_{DC}$. These observations are consistent with the dependence of $E_0$ shown in Fig.~\ref{fig:delta_EGMol_allcases} and of the heating rate demonstrated by molecular dynamic simulations\cite{ryjkov05,tarnas13} with unbounded system. Regarding the influence of $U_{DC}$,  Fig~\ref{fig:main_results} shows  that as $U_{RF}$ is lowered, a high detection efficiency is still achieved for the lowest values of $U_{DC}$. For a finite size system as a cold ion cloud, Eq.~\ref{eq:aspect} shows that decreasing $U_{DC}$ increases the cloud length, suggesting that a lower heating rate and a lower ion density can be compensated by an increased interaction length. The role of the cloud length is more visible on Fig.~\ref{fig:main_results_RandR/L} where the same results are plotted  against the mean value of the measured half-length $L$ of the ion cloud just before the particle injection, $\langle L \rangle$. This representation confirms our previous interpretation that longer clouds gives a higher detection rate for the lowest $U_{RF}$ values.  As the RF-heating is lowered with lower $U_{RF}$, we can conclude that an efficient detection relies on the value of the energy transferred to the cloud by the SUI, that must be sufficient to trigger RF-heating. A lower heating-rate can be compensated  by an increased interaction length of the SUI with the cloud which is independently controlled by the trapping along the axis $U_{DC}$. 




\section{Conclusion}

In the present manuscript we have described the working principle of a new charged particle detector, which is based on the observation of the fluorescence of a cold ion-cloud perturbed by the crossing of a large mass projectile. This detection method is non-destructive and has no upper limit for the mass range, allowing to directly detect giant molecules and (corona-)viruses. We identified the parameters controlling the detection efficiency for a chosen mass  of $10^{6}$ amu. The corresponding experimental set-up is under construction, and combines a molecular source by electro-spray ionisation with the above described linear trap in a differentially pumped vacuum set-up of 1 meter length. The very good control of all parameters makes this set-up interesting also for fundamental explorations. As the radio-frequency heating signal amplification works only for a sufficient energy transferred from the projectile to the cloud, the signal detection can be used to study the experimental conditions that favour this transfer. This will offer an experimental platform to study the stopping power of finite-size one-component plasma, in a regime of large plasma parameter that are out-of-reach of conventional neutral plasma. A better comprehension of the rf-heating mechanism will be an asset for the control of the detection rate\cite{poindron01}.

\begin{acknowledgments}
The authors acknowledge fruitful discussions with Aurika Janulyte. This work benefited from access to the HPC resources of Aix-Marseille Université financed by the project Equip@Meso (Grant No. ANR-10-EQPX-29-01) of the program “Investissements d’Avenir,” supervised by the Agence Nationale de la Recherche. This work has been financially  supported by SATT Sud-Est.
\end{acknowledgments}

\section*{Data Availability statement}
The data that support the findings of this study are available from the corresponding author upon reasonable request.

\section*{Appendix}
The definition of the parameters used in the molecular dynamic simulations is given in table~\ref{tab_def}
\begin{table}[htp]
\caption{List of the parameters used in the molecular dynamic simulations}
\begin{center}
\begin{tabular}{|c|l|}
\hline
$r_0$ & inner radius of the trap \\
$U_{st}$ & static voltage applied to the quadrupole  rods \\
$U_{RF}$ & amplitude of the oscillating voltage applied to the quadrupole rods \\
$U_{DC}$ & voltage difference between the outer and the inner segments of the quadrupole rods\\
$\Omega$ &  frequency of the oscillating voltage $\times 2\pi $\\
$q_x$ & Mathieu parameter coming from the equation of motion of a single ion in the trap \\
$\omega_z$ & oscillation frequency of a Ca$^+$ ion in the potential created along the trap axis $\times 2\pi $  \\
$m_{SUI}$ & mass of the heavy molecule crossing the ion cloud. \\
$Q$  & charge of the heavy molecule (SUI) crossing the ion cloud. \\
$\omega_L$  & laser frequency $\times 2\pi $  \\
$\vec{k}_L$ & laser photon momentum \\
$\omega_0$  & atomic transition frequency $\times 2\pi $  \\
$\delta$ & detuning $\omega_L-\omega_0$ \\
$\Omega_r$  & Rabi frequency of the atom-laser coupling  $\times 2\pi $  \\
$\Gamma$  & spontaneous emission rate of the atom excited state \\
$\omega_x$  &  oscillation frequency of a Ca$^+$ ion in the pseudo-potential  perpendicular to the trap axis created by the time oscillating field $\times 2\pi $  \\
$\omega_r$  &  oscillation frequency of a Ca$^+$ ion in the total static pseudo-potential  perpendicular to the trap axis  $\times 2\pi $  \\
$N$ & number of ions in the cloud \\
$R$  & radius of the cold ion cloud  \\
$L$  & half-length of the cold ion cloud  \\
$T_b$ & thermal bath temperature \\
$T$ & cloud temperature \\
$E_0$ & kinetic energy of the SUI heavy molecule when it reaches the center of the trap\\
\hline
\end{tabular}
\end{center}
\label{tab_def}
\end{table}%

\medskip
\section*{References}

 \bibliographystyle{unsrt}

\begin{thebibliography}{}

\end{thebibliography}


\begin{thebibliography}{10}

\bibitem{shinholt_frequency_2014}
Deven~L. Shinholt, Staci~N. Anthony, Andrew~W. Alexander, Benjamin~E. Draper,
  and Martin~F. Jarrold.
\newblock A frequency and amplitude scanned quadrupole mass filter for the
  analysis of high \textit{m} / \textit{z} ions.
\newblock {\em Review of Scientific Instruments}, 85(11):113109, November 2014.

\bibitem{keifer_charge_2014}
David~Z. Keifer, Elizabeth~E. Pierson, Joanna~A. Hogan, Gregory~J. Bedwell,
  Peter~E. Prevelige, and Martin~F. Jarrold.
\newblock Charge detection mass spectrometry of bacteriophage {P22} procapsid
  distributions above 20 {MDa}: {CDMS} of bacteriophage {P22} procapsid
  distributions above 20 {MDa}.
\newblock {\em Rapid Communications in Mass Spectrometry}, 28(5):483--488,
  March 2014.

\bibitem{blaum2006}
Klaus Blaum.
\newblock High-accuracy mass spectrometry with stored ions.
\newblock {\em Physics Reports}, 425(1):1--78, March 2006.

\bibitem{toyoda_multi-turn_2003}
Michisato Toyoda, Daisuke Okumura, Morio Ishihara, and Itsuo Katakuse.
\newblock Multi-turn time-of-flight mass spectrometers with electrostatic
  sectors.
\newblock {\em Journal of Mass Spectrometry}, 38(11):1125--1142, November 2003.

\bibitem{frank99}
Matthias Frank, Simon~E. Labov, Garrett Westmacott, and W.~Henry Benner.
\newblock Energy-sensitive cryogenic detectors for high-mass biomolecule mass
  spectrometry.
\newblock {\em Mass Spectrometry Reviews}, 18(3‐4):155--186, 1999.

\bibitem{contino_charge_2013}
Nathan~C. Contino, Elizabeth~E. Pierson, David~Z. Keifer, and Martin~F.
  Jarrold.
\newblock Charge {Detection} {Mass} {Spectrometry} with {Resolved} {Charge}
  {States}.
\newblock {\em Journal of The American Society for Mass Spectrometry},
  24(1):101--108, January 2013.

\bibitem{knoop_detection_2014}
Martina Knoop.
\newblock Chapter 2: {Detection} {Techniques} for {Trapped} {Ions}.
\newblock In {\em Physics with {Trapped} {Charged} {Particles}: {Lectures} from
  the {Les} {Houches} {Winter} {School}}, pages 25--42. World Scientific, 2014.

\bibitem{twerenbold01}
Damian Twerenbold, Daniel Gerber, Dominique Gritti, Yvan Gonin, Alexandre
  Netuschill, Frédéric Rossel, Dominique Schenker, and Jean-Luc Vuilleumier.
\newblock Single molecule detector for mass spectrometry with mass independent
  detection efficiency.
\newblock {\em PROTEOMICS}, 1(1):66--69, 2001.

\bibitem{takahashi11}
N~Takahashi, S~Hosokawa, M~Saito, and Y~Haruyama.
\newblock Measurement of absolute detection efficiencies of a microchannel
  plate using the charge transfer reaction.
\newblock {\em Physica Scripta}, T144:014057, jun 2011.

\bibitem{gilmore00}
I.S Gilmore and M.P Seah.
\newblock Ion detection efficiency in sims:: Dependencies on energy, mass and
  composition for microchannel plates used in mass spectrometry.
\newblock {\em International Journal of Mass Spectrometry}, 202(1):217 -- 229,
  2000.

\bibitem{fraser02}
G.W. Fraser.
\newblock The ion detection efficiency of microchannel plates (mcps).
\newblock {\em International Journal of Mass Spectrometry}, 215(1):13 -- 30,
  2002.
\newblock Detectors and the Measurement of Mass Spectra.

\bibitem{twerenbold96b}
Damian Twerenbold, Jean‐Luc Vuilleumier, Daniel Gerber, Almut Tadsen, Ben
  van~den Brandt, and Patrick~M. Gillevet.
\newblock Detection of single macromolecules using a cryogenic particle
  detector coupled to a biopolymer mass spectrometer.
\newblock {\em Applied Physics Letters}, 68(24):3503--3505, 1996.

\bibitem{twerenbold96}
Damian Twerenbold.
\newblock Cryogenic particle detectors.
\newblock {\em Reports on Progress in Physics}, 59(3):349--426, mar 1996.

\bibitem{keifer17}
David~Z. Keifer and Martin~F. Jarrold.
\newblock Single-molecule mass spectrometry.
\newblock {\em Mass Spectrometry Reviews}, 36(6):715--733, 2017.

\bibitem{tito00}
Mark~A. Tito, Kaspar Tars, Karin Valegard, Janos Hajdu, and Carol~V. Robinson.
\newblock Electrospray time-of-flight mass spectrometry of the intact ms2 virus
  capsid.
\newblock {\em Journal of the American Chemical Society}, 122(14):3550--3551,
  2000.

\bibitem{todd20}
Aaron~R. Todd, Andrew~W. Alexander, and Martin~F. Jarrold.
\newblock Implementation of a charge-sensitive amplifier without a feedback
  resistor for charge detection mass spectrometry reduces noise and enables
  detection of individual ions carrying a single charge.
\newblock {\em Journal of the American Society for Mass Spectrometry},
  31(1):146--154, 2020.
\newblock PMID: 32881508.

\bibitem{pat_giantmol}
C.~Champenois, C.~Dedonder-Lardeux, C.~Jouvet, L.~Hilico, M.~Knoop, and
  J.~Pedregosa.
\newblock Non-destructive detection method of charged particles without mass
  limitation.
  \newblock {\em European Patent} 14306498, registered on septembre, 26, 2014

\bibitem{pedregosa10a}
J.~Pedregosa, C.~Champenois, M.~Houssin, and M.~Knoop.
\newblock Anharmonic contributions in real rf linear quadrupole traps.
\newblock {\em International Journal of Mass Spectrometry}, 290(2-3):100 --
  105, 2010.

\bibitem{pedregosa15}
Jofre Pedregosa-Gutierrez, Caroline Champenois, Marius~Romuald Kamsap, and
  Martina Knoop.
\newblock Ion transport in macroscopic \{RF\} linear traps.
\newblock {\em International Journal of Mass Spectrometry}, 381-382:33--40, May
  2015.

\bibitem{kjaergaard00}
N.~Kj{\ae}rgaard, L.~Hornek{\ae}r, A.M. Thommesen, Z.~Videsen, and M.~Drewsen.
\newblock Isotope selective loading of an ion trap using resonance-enhanced
  two-photon ionization.
\newblock {\em Appl. Phys. B}, 71:207--210, 2000.

\bibitem{knoop04}
M.~Knoop, C.~Champenois, G.~Hagel, M.~Houssin, C.~Lisowski, M.~Vedel, and
  F.~Vedel.
\newblock Metastable level lifetimes from electron-shelving measurements with
  ion clouds and single ions.
\newblock {\em Eur. Phys. J. D}, 29:163--171, 2004.

\bibitem{herskind09}
P~F Herskind, A~Dantan, M~Albert, J~P Marler, and M~Drewsen.
\newblock Positioning of the rf potential minimum line of a linear paul trap
  with micrometer precision.
\newblock {\em Journal of Physics B: Atomic, Molecular and Optical Physics},
  42(15):154008, 2009.

\bibitem{dehmelt67}
H.G. Dehmelt.
\newblock Radiofrequency spectroscopy of stored ions {I}: storage.
\newblock {\em Advances in Atomic and Molecular Physics}, 3:53--72, 1967.

\bibitem{drewsen00}
M.~Drewsen and A.~Br{\o}ner.
\newblock Harmonic linear {P}aul trap: {S}tability diagram and effective
  potentials.
\newblock {\em Phys. Rev. A}, 62:045401, 2000.

\bibitem{drakoudis06}
A.~Drakoudis, M.~Söllner, and G.~Werth.
\newblock Instabilities of ion motion in a linear paul trap.
\newblock {\em International Journal of Mass Spectrometry}, 252(1):61 -- 68,
  2006.

\bibitem{turner87}
Leaf Turner.
\newblock Collective effects on equilibria of trapped charged plasmas.
\newblock {\em Phys. Fluids}, 30:3196, 1987.

\bibitem{hornekaer02}
L.~Hornek{\ae}r and M.~Drewsen.
\newblock Formation process of large ion coulomb crystals in linear {P}aul
  traps.
\newblock {\em Phys. Rev. A}, 66(1):013412, Jul 2002.

\bibitem{champenois09}
C~Champenois.
\newblock About the dynamics and thermodynamics of trapped ions.
\newblock {\em Journal of Physics B: Atomic, Molecular and Optical Physics},
  42(15):154002 (9pp), 2009.

\bibitem{Major05}
Fouad~G. Major, Viorica~N. Gheorghe, and G\"unther Werth.
\newblock {\em Charged Particle Traps}, volume~37 of {\em Springer Series on
  Atomic, Optical, and Plasma Physics}.
\newblock Springer Berlin Heidelberg, 2005.
\newblock 10.1007/3-540-26576-7\_1.

\bibitem{skeel02}
Robert~D. Skeel and Jesus~A. Izaguirre.
\newblock An impulse integrator for langevin dynamics.
\newblock {\em Molecular Physics}, 100(24):3885--3891, 2002.

\bibitem{pedregosa20}
J~Pedregosa-Gutierrez and M~Mukherjee.
\newblock Defect generation and dynamics during quenching in finite size
  homogeneous ion chains.
\newblock {\em New Journal of Physics}, 22(7):073044, July 2020.

\bibitem{marciante10}
M.~Marciante, C.~Champenois, A.~Calisti, J.~Pedregosa-Gutierrez, and M.~Knoop.
\newblock Ion dynamics in a linear radio-frequency trap with a single cooling
  laser.
\newblock {\em Phys. Rev. A}, 82(3):033406, Sep 2010.

\bibitem{simion}
Scientific Instrument Services, Inc.; SIMION 8.1; http://www.simion.com.

\bibitem{sillitoe17}
Nicolas Sillitoe, Jean-Philippe Karr, Johannes Heinrich, Thomas Louvradoux,
  Albane Douillet, and Laurent Hilico.
\newblock $\overline{{\rm H}}^{+}$ {Sympathetic} {Cooling} {Simulations} with a
  {Variable} {Time} {Step}.
\newblock JPS Conf. Proc. {\bf 18}, 011014 (2017).
\newblock https://doi.org/10.7566/JPSCP.18.011014

\bibitem{champenois16}
C.~Champenois.
\newblock {\em Trapped Charged Particles, Laser Cooling Techniques Applicable
  to Trapped Ions}, page 117.
\newblock World Scientific, 2016.

\bibitem{prestage91b}
J.~D. Prestage, A.~Williams, L.~Maleki, M.~J. Djomehri, and E.~Harabetian.
\newblock Dynamics of charged particles in a {P}aul radio-frequency quadrupole
  trap.
\newblock {\em Phys. Rev. Lett.}, 66(23):2964--2967, 1991.

\bibitem{schiffer00}
J.P. Schiffer, M.~Drewsen, J.S. Hangst, and L.~Hornek{\ae}r.
\newblock Temperature, ordering,and equilibrium with time-dependent confining
  force.
\newblock {\em PNAS}, 97:10697, 2000.

\bibitem{drewsen98}
M.~Drewsen, C.~Brodersen, L.~Hornek\ae{}r, J.~S. Hangst, and J.~P. Schiffer.
\newblock Large ion crystals in a linear {P}aul trap.
\newblock {\em Phys. Rev. Lett.}, 81(14):2878--2881, 1998.

\bibitem{lett89}
P.D. Lett, W.D. Phillips, S.L. Rolston, C.E. Tanner, R.N. Watts, and
  C.I.Westbrook.
\newblock Optical molasses.
\newblock {\em J. Opt. Soc. Am. B}, 6(11):2084, 1989.

\bibitem{alheit95}
R.~Alheit, C.~Henning, R.~Morgenstern, F.~Vedel, and G.~Werth.
\newblock Observation of instabilities in a {P}aul trap with higher-order
  anharmonicities.
\newblock {\em Appl. Phys. B}, 61:277--283, 1995.

\bibitem{alheit96}
R.~Alheit, S.~Kleineidam, F.~Vedel, M.~Vedel, and G.~Werth.
\newblock Higher order non-linear resonances in a {P}aul trap.
\newblock {\em Int.J. Mass Spectrom. Ion Processes}, 154:155--169, 1996.

\bibitem{ryjkov05}
Vladimir~L. Ryjkov, XianZhen Zhao, and Hans~A. Schuessler.
\newblock Simulations of the rf heating rates in a linear quadrupole ion trap.
\newblock {\em Phys. Rev. A}, 71(3):033414, Mar 2005.

\bibitem{tarnas13}
J.~D. Tarnas, Y.~S. Nam, and R.~Blümel.
\newblock Universal heating curve of damped {Coulomb} plasmas in a {Paul} trap.
\newblock {\em Physical Review A}, 88(4), October 2013.

\bibitem{zwicknagel_stopping_1999}
Günter Zwicknagel, Christian Toepffer, and Paul-Gerhard Reinhard.
\newblock Stopping of heavy ions in plasmas at strong coupling.
\newblock {\em Physics reports}, 309(3):117--208, 1999.

\bibitem{bussmann06b}
M.~Bussmann, U.~Schramm, and D.~Habs.
\newblock Simulating the stopping dynamics of highly charged ions in an
  ultra-cold, strongly coupled plasma.
\newblock {\em Hyperfine Interactions}, 173(1-3):27--34, November 2006.

\bibitem{poindron01}
A.~Poindron {\it et. al}.
\newblock Heating of Coulomb Crystals in Linear RF Traps
\newblock to be published, 2021.

\end{thebibliography}

\end{document}